\documentclass[aps,pre,superscriptaddress,unsortedaddress,twocolumn]{revtex4}

\usepackage{amsfonts}

\usepackage{graphicx}
\usepackage{epstopdf}
\usepackage{amsmath}
\usepackage{amssymb}
\usepackage{verbatim}
\begin{document}

\title{Brownian Carnot engine} 

\author{Ignacio A. Mart\'inez$^\star$}
\affiliation{ICFO $-$ Institut de Ci\`encies Fot\`oniques, Mediterranean Technology Park,  Av. Carl Friedrich Gauss, 3, 08860, Castelldefels (Barcelona), Spain.}
\affiliation{Laboratoire de Physique, \'Ecole Normale Sup\'erieure, CNRS UMR5672 46 All\'ee d'Italie, 69364 Lyon, France.}
\author{\'Edgar Rold\'an\footnote{These authors contributed equally to this work}}
\affiliation{ICFO $-$ Institut de Ci\`encies Fot\`oniques, Mediterranean Technology Park,  Av. Carl Friedrich Gauss, 3, 08860, Castelldefels (Barcelona), Spain.}
\affiliation{ Max Planck Institute for the Physics of Complex Systems,  N{\"o}thnitzer Str. 38, 01187 Dresden, Germany.}
\affiliation{GISC - Grupo Interdisciplinar de Sistemas Complejos, Madrid, Spain.}
\author{Luis Dinis}
\affiliation{Departamento de Fisica At\'omica, Molecular y Nuclear, Universidad Complutense Madrid, 28040 Madrid, Spain}
\affiliation{GISC - Grupo Interdisciplinar de Sistemas Complejos, Madrid, Spain.}
\author{Dmitri Petrov}
\affiliation{ICFO $-$ Institut de Ci\`encies Fot\`oniques, Mediterranean Technology Park,  Av. Carl Friedrich Gauss, 3, 08860, Castelldefels (Barcelona), Spain.}
\author{J. M. R. Parrondo}
\affiliation{Departamento de Fisica At\'omica, Molecular y Nuclear, Universidad Complutense Madrid, 28040 Madrid, Spain}
\affiliation{GISC - Grupo Interdisciplinar de Sistemas Complejos, Madrid, Spain.}
\author{Ra\'ul A. Rica\footnote{Correspondence should be addressed to Ignacio A. Mart\'inez (Email: \href{mailto:martinez.ignacio@ens-lyon.fr}{martinez.ignacio@ens-lyon.fr}), \'Edgar Rold\'an (Email: \href{mailto:edgar@pks.mpg.de}{edgar@pks.mpg.de}) and Ra\'ul A. Rica (Email: \href{mailto:rul@ugr.es}{rul@ugr.es}).}}
\affiliation{ICFO $-$ Institut de Ci\`encies Fot\`oniques, Mediterranean Technology Park,  Av. Carl Friedrich Gauss, 3, 08860, Castelldefels (Barcelona), Spain.}

\maketitle

%\pacs{
%05.70.Ln,  %	Nonequilibrium and irreversible thermodynamics 
%05.40.-a   %    Fluctuation phenomena, random processes, noise, and Brownian motion
%05.70.-a   %    Entropy thermodynamics
%}

\textbf{The Carnot cycle imposes a fundamental upper limit to the  efficiency of a macroscopic motor operating between two thermal baths~\cite{carnot1872reflexions}. 
However, this bound needs to be reinterpreted at microscopic scales, where molecular bio-motors \cite{Howard2001} and some  artificial micro-engines \cite{blickle2012realization,Roldan2014universal,koski2014experimental}  operate.
As described by stochastic thermodynamics \cite{sekimoto2010stochastic,seifert2012stochastic},  energy transfers in microscopic systems are random and thermal fluctuations induce transient decreases of entropy, allowing for possible violations of the Carnot limit~\cite{wang2002experimental}. 
Despite  its potential relevance for the development of a thermodynamics of small systems, an experimental study of microscopic Carnot engines is still lacking. Here we report on an experimental realization of a Carnot engine with a single optically trapped Brownian particle as working substance. We present an exhaustive study of the energetics of the engine and analyze the fluctuations of the finite-time efficiency, showing that the Carnot bound can be surpassed for a small number of non-equilibrium cycles. As its macroscopic counterpart, the energetics of our Carnot device exhibits basic properties that one would expect to observe in any microscopic energy transducer operating with baths at different temperatures~\cite{polettini2015efficiency,
verley2014universal,gingrich2014efficiency}. Our results characterize the sources of irreversibility in the engine and the statistical properties of the efficiency ---an insight that could inspire novel strategies
in the design of efficient  nano-motors.
} 

%Thermodynamics was developed in the XIX century using the Carnot engine as a building block. Carnot realized that the maximum efficiency $\eta_{\rm C}$ attainable by a thermal engine working between  two reservoirs at different temperatures (hot, $T_{\rm h}$, and cold, $T_{\rm c}$) is a universal function of the temperatures $T_{\rm h}>T_{\rm c}$ \cite{carnot1872reflexions}. This universality was used by Kelvin to define a fundamental scale of  temperature, obeying $\eta_{\rm C}=1-T_{\rm c}/T_{\rm h}$, and by Clausius to introduce the concept of  entropy. Years after, the Carnot cycle also played a prominent role in the development of finite time thermodynamics, when Curzon and Ahlborn obtained the efficiency at maximum power of an irreversible engine~\cite{curzon1975efficiency}. 

The Carnot cycle consists of two isothermal processes, where the working substance is respectively in contact with thermal baths at different temperatures $T_{\rm h}$ and $T_{\rm c}$, connected by two adiabatic processes, where the substance is isolated and heat is not delivered nor absorbed. 
An external parameter is changed in such a way that the whole cycle is carried out reversibly. 
Following this scheme, one could devise a progressing miniaturization of a Carnot engine and eventually reproduce the cycle with a single Brownian particle. In fact, a variety of thermodynamic processes and even a complete Stirling cycle have been already implemented in the mesoscale using micro manipulation techniques \cite{toyabe2010experimental,blickle2012realization,Roldan2014universal,koski2014experimental,roldan2014measuring,martinez2014adiabatic}. Interestingly, the exchange of energy between the particle and its surrounding environment becomes stochastic at the micro scale and yet one can rigorously define work, heat, and efficiency, within the  framework of the recently developed {\em stochastic thermodynamics} \cite{sekimoto2010stochastic,seifert2012stochastic}.

The experimental realization of a Carnot cycle with a single Brownian particle has remained elusive due to the difficulties of implementing an adiabatic process. In particular, it is not clear how to isolate a particle from the surrounding fluid~\cite{crooks2007work}. A more feasible strategy is to simultaneously change the temperature and the external parameter keeping constant the Shannon entropy of the particle. However,  the necessary fine tuning of the temperature is an experimental challenge as well. Here we construct a Brownian Carnot engine putting forward an experimental technique that allows a precise control of both the effective temperature and the accesible volume of a single microscopic particle (See Methods and~\cite{martinez2013effective,mestres2014realization, berut2014energy}). We use a particle with an inherent electric charge and apply a noisy electrostatic force that mimics a thermal bath. In this way, we can achieve temperatures  ranging  from room temperature (no electrostatic force) up to hundreds or even thousands of Kelvins, far above the boiling point of water.

The working substance of our engine is a single optically trapped colloidal particle immersed in water \cite{martinez2014adiabatic}. 
For small displacements $x$ from the trap equilibrium position, the optical potential is harmonic, $U(x,t)=\kappa x(t)^2/2$, with stiffness $\kappa$.
The Hamiltonian or total energy of the particle is $H=\kappa x^2/2+p^2/(2m)$,  $p=m \frac{\text{d} x}{\text{d}t}$ being the linear momentum of the particle and $m$ the mass of the particle. The conjugated force for the external parameter $\kappa$ is $F_\kappa(t) \equiv  \partial H/\partial \kappa=  x^2(t)/2$. As a result, the work necessary to implement a change $\text{d}\kappa$ in the external parameter, ${\rm d} W(t)=F_\kappa(t){\rm d}\kappa$, and the heat or energy transfer from the thermal bath to the particle, $ \text{d} Q(t)=\text{d}H(t)-\text{d} W(t)$, are  fluctuating quantities.

\begin{figure*}
\centering
\includegraphics[width= 14cm]{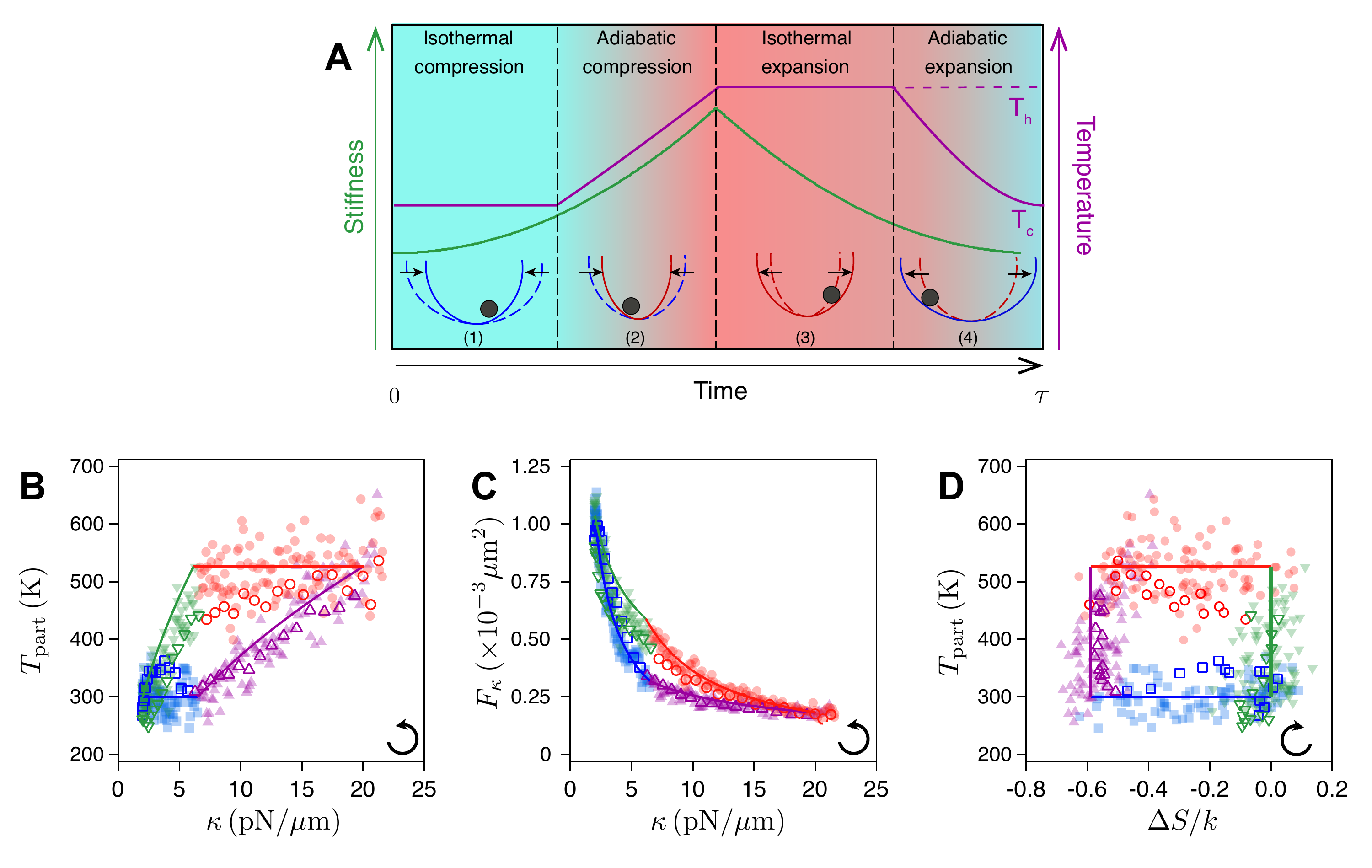}
\caption[The Brownian Carnot engine.] {\small\label{fig:scheme} {\bfseries The Brownian Carnot engine.} \textbf{(A)} Time evolution of the experimental protocol.  \textbf{(B-D)} Thermodynamic diagrams of the engine:   (1) Isothermal compression (blue); (2) Adiabatic compression 
 (magenta); (3) Isothermal expansion (red);  (4) Adiabatic expansion (green). Solid lines are the analytical values in the quasistatic limit. Filled symbols are obtained from ensemble averages over cycles of duration $\tau=200\,\rm ms$ while open symbols are obtained for $\tau=30\,\rm ms$. 
The black arrow indicates the direction of the operation of the engine.  \textbf{(B)} $T_{\rm part}-\kappa$ diagram.  
 \textbf{(C)} Clapeyron diagram. The area within the cycle is equal to the mean work obtained during the cycle. \textbf{(D)} $T_{\rm part}$--$S$ diagram. The entropy changes only in the isothermal steps.} % The entropy change is calculated as the difference between the Shannon entropy of the full phase space at any time $t$ and its value at the beginning of the cycle.}
\end{figure*}

The Carnot cycle is implemented by modifying the stiffness $\kappa$ and the environment temperature $T$ (Figs.~\ref{fig:scheme}A-B) and consists of two isothermal processes ($T$ is kept constant and $\kappa$ changes, blue and red curves in Fig.~\ref{fig:scheme}B) and two adiabatic processes ($T$ and $\kappa$ change keeping $T^2/\kappa$ constant \cite{martinez2014adiabatic}, green and magenta curves in Fig.~\ref{fig:scheme}B). 
We measure different thermodynamic quantities (temperature, stiffness, heat, work and Shannon entropy, see Methods) under both equilibrium and non-equilibrium driving (Figs.~\ref{fig:scheme}B-D). 
The effective temperature of the particle is obtained from the average potential energy, $T_{\rm part}(t)\equiv\kappa(t)\langle x(t)^2\rangle/k$, and can differ from the environment temperature $T$ for non quasistatic protocols.
The $T_{\rm part}-\kappa$ diagram of the engine (Fig.~\ref{fig:scheme}B) shows larger fluctuations in the quasistatic equilibrium protocol, because the average is taken over a smaller number of cycles. In the non-equilibrium protocol, the most irreversible steps are the expansions, where the particle remains colder (i.e. more confined~\cite{gieseler2012subkelvin}) than the environment. As in a macroscopic gas, the expansion is dominated by an entropic force, namely, the tendency of the gas to fill the available space. In the case of the single Brownian particle, the expansion is driven by thermal fluctuations that allow the particle to move farther away from the center of the trap. On the other hand, the compression is driven by the trap confining force, which allows the particle to react more rapidly and to follow the equilibrium temperature even in fast cycles in the adiabatic compression. In the isothermal compression, however, we observe a fast initial increase of the temperature of the particle due to the increase of the stiffness. The $F_{\kappa} - \kappa$ diagram (Fig.~\ref{fig:scheme}C) resembles Clapeyron pressure vs volume diagram of a Carnot cycle performed with an ideal gas~\cite{feynman}. 
The  $T_{\rm part}-S$ diagram of the particle (Fig.~\ref{fig:scheme}D) is a rectangle where all the entropy changes in the system occur in the two isothermal steps. This diagram also gives information about the nature of the irreversibility for a fast driving (open symbols): the effective temperature of the particle in the isothermal processes suggests the presence of an irreversible flow of energy between the reservoir and the particle, resembling the endo-reversible  engine introduced by Curzon and Ahlborn \cite{curzon1975efficiency,Ouerdane:2015dd}

%Taken all together, the thermodynamic diagrams under quasistatic driving (Figs.~\ref{fig:scheme}B-D) are equivalent to those for a single particle ideal gas in a Carnot cycle~\cite{feynman}.

 \begin{figure}
\centering
\includegraphics[width= 9cm]{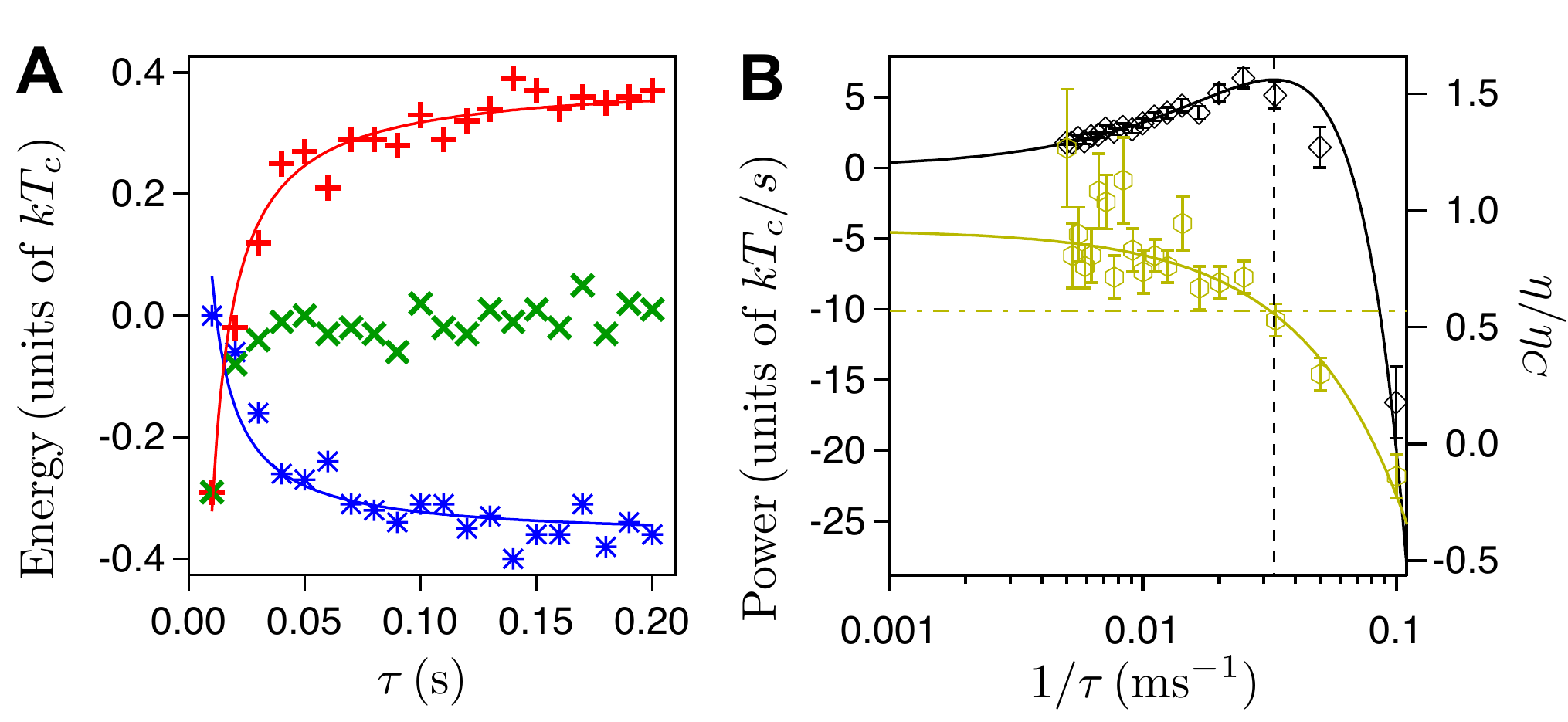}
\caption[Energetics of the Brownian Carnot engine.] {\small\label{fig:energetics} {\bfseries Energetics of the Brownian Carnot engine.} \textbf{(A)} Ensemble averages of stochastic work ($\langle W_\tau\rangle$, blue stars)  and heat ($\langle Q_\tau\rangle$, red pluses) transferred in one cycle as a function of the cycle duration. Green crosses are the average total energy change of the working substance $\langle \Delta H_\tau\rangle$. Thin lines are fits to Sekimoto-Sasa law, $A+B/\tau$. \textbf{(B)} Power output $P_\tau=-\langle W_\tau\rangle /\tau$  (black diamonds, left axis) and long-term efficiency $\eta_\tau$ (yellow hexagons, right axis) as a function of the inverse of the cycle time. The black  curve is a fit $P_\tau= (\langle W_{\infty}\rangle + \Sigma_{\rm ss}/\tau)/\tau $, yielding  $\langle W_{\infty}\rangle  = (- 0.38\pm 0.01) kT_c$  and $\Sigma_{\rm ss} = (5.7\pm 0.3)kT_c \,\rm ms$ with a reduced chi-square of $\chi^2_{\rm red}=1.08$. The solid yellow line is a fit to $\eta_\tau = (\eta_C+\tau_W/\tau)/(1+\tau_Q/\tau)$, which yields $\eta_{\infty} = (0.92\pm 0.06)\eta_C$, $\tau_W=(-11\pm 2)\,\rm ms$, $\tau_Q=(-0.6\pm 6.0)\,\rm ms$ with $\chi^2_{\rm red}= 0.76$.  Yellow dash-dot line is the Curzon-Alborn efficiency $\eta_{\rm CA}= 1-\sqrt{T_c/T_h} = 0.25=0.57\eta_C$, which is in excellent agreement with the location of the maximum power (vertical black dashed line). Ensemble averages are done over $50\,\rm s$  and error bars are obtained with a statistical significance of $90\%$. }
\end{figure}
During a cycle of duration $\tau$, the working substance of the engine exchanges heat with the different thermal baths it is put in contact with, and under appropriate conditions it is able to extract work. We call $W_{\tau}$ and $Q_{\tau}$ the work exerted on the particle and the heat transferred from the environment to the particle along a cycle, respectively. The exchanged heat equals to $Q_\tau = \Delta H_\tau - W_\tau$. Both work and heat along the whole cycle (Fig.~\ref{fig:energetics}A) converge to their quasistatic averages $\langle \cdot_{\infty}\rangle$ following Sekimoto and Sasa's law $\langle W_\tau \rangle =\langle W_{\infty}\rangle + \Sigma_{\rm ss} / \tau$~\cite{sekisasa1997complementarity}.  Here, $\langle W_{\infty}\rangle$ is the quasistatic value of the work done per cycle and the term $ \Sigma_{\rm ss} / \tau$ accounts for the (positive) dissipation, which decays to zero like $1/\tau$~\cite{bonanza2014optimal}. In the case of the average heat per cycle, $\langle Q_\tau\rangle$, we find the dissipative term is negative, i.e., $\langle Q_\tau\rangle =\langle Q_{\infty}\rangle - \Sigma_{\rm ss} / \tau$ with $\Sigma_{\rm ss}>0$. 
 
 \begin{figure}
\centering
\includegraphics[width= 8.5cm]{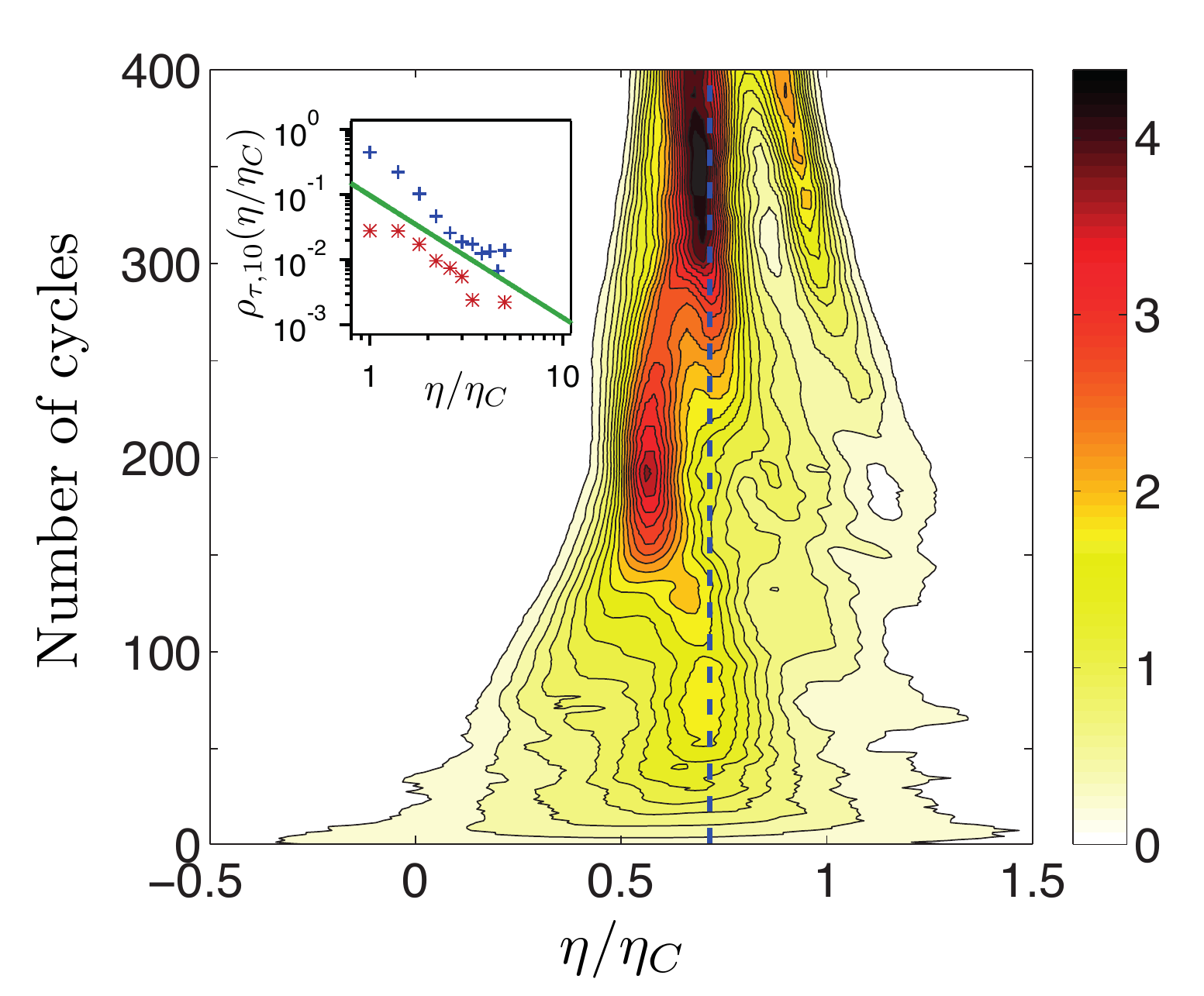}
\caption[Efficiency fluctuations at maximum power.] {\small\label{fig:efficiencies} {\bfseries Efficiency fluctuations at maximum power.} Contour plot of the probability density function of the efficiency $\rho_{\tau=40\,\text{ms},i} (\eta)$ computed summing over $i=1$ to $400$ cycles (left axis). The  long-term efficiency (averaged over $\tau_{\rm exp} = 50\,\rm s$) is shown with a vertical blue dashed line. Super Carnot efficiencies appear even far from quasistatic driving. {\em Inset}: Tails of the distribution for $\rho_{\tau=40\,\text{ms},10} (\eta)$ (blue squares, positive tail; red circles, negative tail). The green line is a fit to a power-law to all the data shown, whose exponent is $\gamma = (-1.9\pm 0.3)$.}
\end{figure}
To quantify the performance of the engine, we analyze its power output and efficiency. First, we measure the power output as the mean total work exchanged during a cycle divided by the total duration of the cycle (Fig.~\ref{fig:energetics}B), $P_\tau=-\langle W_\tau\rangle/\tau$. For $\tau=10\,\rm ms$, $\langle W_\tau \rangle$ is positive, the particle behaves as a heat pump and the power is negative. For larger values of $\tau$ the power increases, becoming positive, and eventually reaches a maximum value  $P_{\rm max} = 6.34\, kT_{\rm c}/s$. Above that maximum, $P_\tau$ decreases monotonically when increasing the cycle length. The data of $P_\tau$ vs $\tau$ fits well to the expected law $P_\tau=-(\langle W_{\infty}\rangle + \Sigma_{\rm ss} / \tau)/\tau$. 
The efficiency is given by the ratio between the extracted work and the input of heat, which is usually considered as the heat flowing from the hot thermal bath to the system.
In our experiment, however, there is a non zero  fluctuating heat in the adiabatic steps, which must be taken into account in the definition of the stochastic efficiency of the engine during a finite number of cycles. Here  we will consider this heat as input (see Methods for alternative definitions of the efficiency). We define  $W_{\tau}^{(i)}$ as the sum of the total work exerted on the particle along $i\geq 1$ cycles of duration $\tau$, and $Q_{\alpha,\tau}^{(i)}$ the sum over $i$ cycles of the heat transferred to the particle in the $\alpha-$th subprocess ($\alpha=1,2,3,4$, cf. Fig.~\ref{fig:scheme}).  We therefore introduce the following definition of stochastic efficiency:
\begin{equation}
\eta_{\tau}^{(i)} = \frac{- W_{\tau}^{(i)}}{ Q_{2,\tau}^{(i)} + Q_{3,\tau}^{(i)}+ Q_{4,\tau}^{(i)}}.
\label{eq:stoch_eff_3}
\end{equation}
The long-term efficiency of the motor is given by  $\eta_\tau\equiv\eta_{\tau}^{(i)}$ with $i\to\infty$. In the quasistatic limit, the average heat in the adiabatic processes vanishes yielding $\eta_\infty =  \eta_{\rm C}\equiv 1-T_{\rm c}/T_{\rm h}\simeq 0.43$ (Fig.~\ref{fig:energetics}B). Moreover, the standard efficiency at maximum power, $\eta^\star\simeq (0.25\pm 0.05)$, is in agreement with the Curzon-Ahlborn expression for finite-time cycles $\eta_{\rm{CA}}=1-\sqrt{T_c/T_h}\simeq 0.25$~\cite{curzon1975efficiency,esposito2010efficiency}.

Very recently, much attention has been drawn to the statistical properties of the efficiency of stochastic engines. %In a seminal paper
Using fluctuation theorems, Verley {\em et al.}  proved that the probability distribution function (PDF)  of the efficiency of an autonomous or symmetrically driven engine has a local minimum precisely at the Carnot value $\eta_{\rm C}$~\cite{verley2014unlikely}. For non-symmetric driving protocols, like our Carnot cycle, there are several theoretical predictions concerning the PDF as well as the large deviation function (LDF) of the stochastic efficiency 
\cite{verley2014universal,gingrich2014efficiency}.  In order to test some of these predictions, we measure the PDF $\rho_{\tau,i}(\eta)$  of the stochastic efficiency $\eta_\tau^{(i)}$ (Methods). Close to equilibrium, near the maximum power output of the engine, the distribution is bimodal when summing over several cycles (Fig.~\ref{fig:efficiencies})~\cite{gingrich2014efficiency,polettini2015efficiency}.  Indeed, local maxima of $\rho_{\tau,i}(\eta)$ appear above standard efficiency for large values of $i$. Another universal feature tested here is that the tails of the distribution follow a power-law, $\rho_{\tau,i}(\eta\to\pm\infty)\sim\eta^{-2}$ (inset of Fig.~\ref{fig:efficiencies})~\cite{proesmans2015stochastic,gingrich2014efficiency}. In the Supplementary Information, we discuss in detail and provide further experimental tests of other  universal properties of  the PDF and the LDF of the stochastic efficiency.

We have realized the first Brownian Carnot engine with a single microscopic particle as a working substance which is able to transform the heat transferred from thermal fluctuations into mechanical work, characterizing both its mean behavior and fluctuations. At slow driving, our engine attains the fundamental limit of Carnot efficiency. The maximum power performed by our engine is $\sim250$ larger than that of previous micro-engines~\cite{blickle2012realization} and only one order of magnitude below the power developed by some biological molecular motors such as myosin~\cite{Howard2001}. 
Our results could be exploited in the design of novel biologically-inspired nano engines~\cite{sarikaya2003molecular} or artificial nanorobots~\cite{douglas2012logic}. In vacuum, trapping techniques could benefit from our study of the efficiency fluctuations to build engines capable to outperform Carnot efficiency~\cite{rossnagel2014nanoscale,gieseler2014dynamic,millen2014nanoscale}.

%--------

  \vskip 2em
I.A.M., E.R., D.P. and R.A.R. acknowledge financial support from Fundaci\'o Privada Cellex Barcelona. I.A.M., D.P. and R.A.R. acknowledge financial support from grant NanoMQ (FIS2011-24409. MINECO). I.A.M. acknowledges financial support from  the European Research Council Grant OUTEFLUCOP. E.R., L.D. and J.M.R.P. acknowledge financial support from grant ENFASIS (FIS2011-22644, MINECO) and TerMic (FIS2014-52486-R, MINECO).  We wish to acknowledge the work of Stephen Corcuff at the earliest stage of the project and fruitful discussions with Ricardo Brito. Prof. D. Petrov passed away on 3 February 2014. He impulsed the development of this project while he was the leader of the Optical Tweezers group at ICFO. We mourn the loss of a great colleague and friend.

\section*{SUPPLEMENTARY INFORMATION}
\section{Experimental tests on stochastic efficiency}
\label{sec:tests}

The fluctuations of the efficiency of heat engines have been been characterized in the framework of stochastic thermodynamics \cite{sekimoto200carnot,verley2014unlikely,gingrich2014efficiency,polettini2015efficiency,rana2014single}.  Universal properties of the probability density function (PDF) \cite{polettini2015efficiency} of the efficiency and of its large deviation function (LDF) \cite{verley2014unlikely,gingrich2014efficiency,verley2014universal} have  recently been established from the application of fluctuation theorems to mesoscopic engines. The experimental verification of the majority of these results is however still lacking.

Two of the major theoretical predictions of the efficiency PDF, namely the bimodality of the histogram near the maximum power output \cite{polettini2015efficiency} and the power-law tails \cite{proesmans2015stochastic}, have been tested experimentally in Fig. 3 in the Main Text. In what follows, we review some of the main theoretical predictions for the LDF of the stochastic efficiency and show the experimental test of some of these features in our Carnot micro engine.

In the limit of large observation times, the efficiency distribution can be characterized by its LDF. The LDF of the efficiency fluctuations, $J_{\tau}(\eta)$, describes the asymptotic behaviour of the efficiency PDF  when the efficiency is calculated summing over a large number of cycles
\begin{equation}
\rho(\eta_{\tau}^{(i)}) \simeq e^{-i J_{\tau}(\eta)}\quad,\quad \text{for} \quad i\to\infty\quad,
\label{eq:LDF_definition}
\end{equation}
where the subindex $\tau$ in $J_{\tau}(\eta)$ indicates the duration of the cycle of the engine. Here $\rho(\eta_{\tau}^{(i)})$ is the PDF of the efficiency obtained as the ratio of the cumulative sum of the work $W$ over $i$ cycles over the total heat absorbed ($Q=Q_1+Q_2+Q_3$) summed over $i$ cycles. From Eq.~\eqref{eq:LDF_definition}, the LDF of the efficiency can be estimated as
\begin{equation}
J_{\tau}(\eta)\simeq -\lim_{i\to\infty}\frac{1}{i}\ln\rho(\eta_{\tau}^{(i)}) \quad.
\label{eq:LDF_estimation}
\end{equation}
Introducing the observation time $\tau_{\rm obs}= i\tau$, we can also define the LDF in units of inverse time,
\begin{equation}
\mathcal{J}_{\tau}(\eta)\simeq -\lim_{\tau_{\rm obs}\to\infty}\frac{1}{\tau_{\rm obs}}\ln\rho(\eta_{\tau}^{(\tau_{\rm obs})})\quad.
\label{eq:LDF_estimation2}
\end{equation}

In time-symmetric cycles the efficiency LDF has been found to attain its global maximum at the Carnot value $\eta_C$ \cite{verley2014unlikely}. In other words, Carnot efficiency is the least likely efficiency  $\eta_{\rm min}$ in time-symmetric cycles: $\eta_{\rm min}^{\rm sym}=\eta_C$. For time-asymmetric cycles like our Carnot engine, however, an off-Carnot maximum of the LDF  has been predicted theoretically, $\eta_{\rm min}^{\rm asym}\neq\eta_C$ \cite{gingrich2014efficiency,verley2014universal}. In such a case, the stochastic entropy production $\Delta S_{\rm tot}$ vanishes when averaged over ensembles of trajectories whose efficiency equals to $\eta_C$ and also when averaged over ensembles of trajectories whose efficiency equals to $\eta_{\rm min}$ \cite{gingrich2014efficiency}: $\langle\Delta S_{\rm tot}\rangle_{\eta}=0$ for $\eta=\eta_C,\eta_{\rm min}$.

\begin{figure}
\centering
\includegraphics[width=6cm]{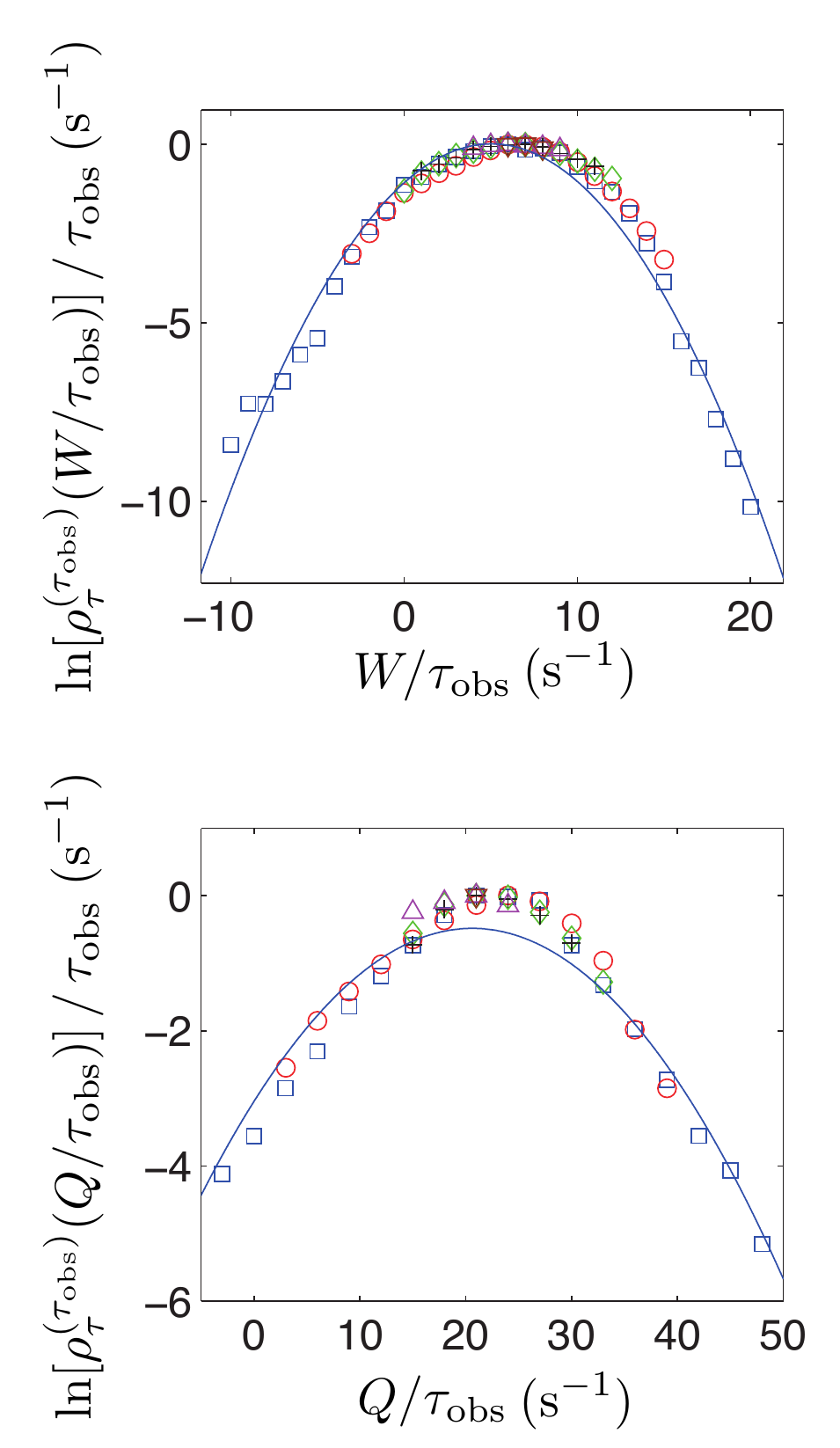}
\caption{\textbf{Work and heat fluctuations for the Carnot cycle of duration $\tau=40\rm ms$}.  Top: Work rate function $\ln [\rho_\tau^{\tau_{\rm obs}}(W/\tau_{\rm obs})]/\tau_{\rm obs}$  as a function of the  work scaled by the observation time. The data is obtained for different observation times $\tau_{\rm obs}$ corresponding to $10$ cycles (blue squares), $20$ cycles (red circles), $40$ cycles (green diamonds), $50$ cycles (black "$+$"), $100$ cycles (magenta up triangles) and $200$ cycles (brown down triangles). Bottom: Rate function of the total absorbed heat measured at the same observation times. Both heat and work are measured in units of $kT_c$, with $T_c=300\,\rm K$ and $k$ Boltzmann's constant. All the distributions are normalized to their maximum value. The solid lines are fits of the $10$-cycle distributions to a Gaussian distribution.  \label{fig:LDFWQ}
%bin size 1 for W 3 for Q
}
\end{figure}

The first theoretical results on stochastic efficiency have been obtained under the assumption of Gaussian work and heat fluctuations \cite{polettini2015efficiency} or when the joint distribution of heat and work is smooth around zero \cite{gingrich2014efficiency}. Figure \ref{fig:LDFWQ} shows that the experimental distributions of the extracted work and the total absorbed heat can be approximatively described as Gaussians. Note that the work distributions at an observation time equivalent to $10$ cycles fit better to a Gaussian distribution than that heat summed over $10$ cycles of the engine. Both work and heat rate functions $\ln[\rho_{\tau}^{(\tau_{\rm obs})}(W/\tau_{\rm obs})]/\tau_{\rm obs}$ and $\ln[\rho_{\tau}^{(\tau_{\rm obs})}(Q/\tau_{\rm obs})]/\tau_{\rm obs}$ collapse to a universal curve for observation times equal or larger than $\sim 20$ cycles (cf. Fig. 2 in \cite{gingrich2014efficiency}).

\begin{figure}
\centering
\includegraphics[width=6cm]{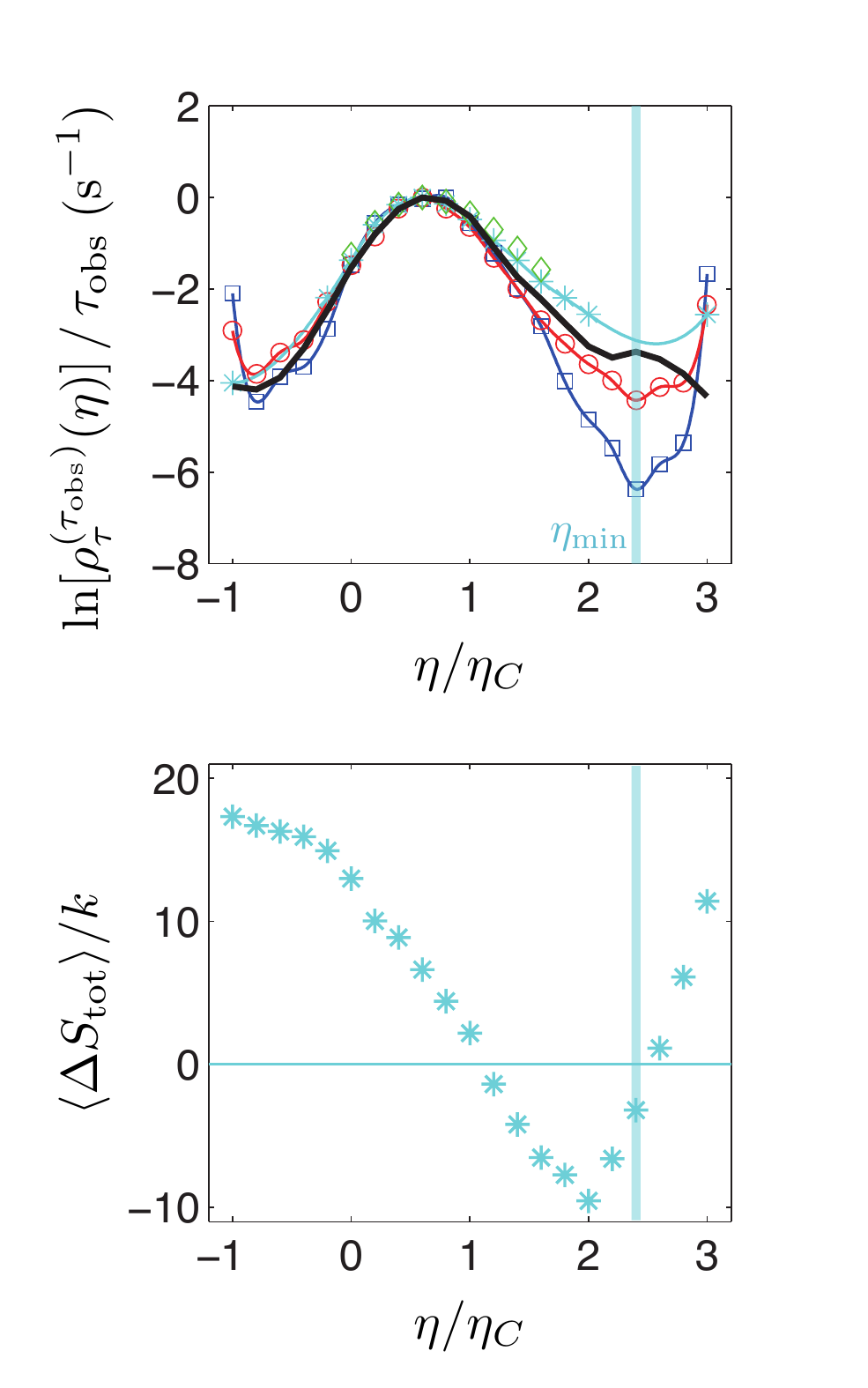}
\caption{\textbf{Efficiency large deviation function and mean entropy production for  the Carnot cycle of duration $\tau=40\rm ms$}.  Top: Rate function of the efficiency  (normalized to the maximum value) for different observation times corresponding to $10$ cycles (blue squares), $20$ cycles (red circles), $30$ cycles (cyan stars) and $40$ cycles (green diamonds). The data is obtained using a regular binning from $-\eta_C$ to $3\eta_C$ with bin size $0.2\eta_C$. The black line is the efficiency LDF calculated using the method described in Sec. \ref{sec:method} and the solid lines connecting the symbols are obtained with a spline interpolation. Bottom: Mean entropy production as a function of the efficiency for an observation time of $30$ cycles.  Mean entropy production vanishes at $\eta\simeq \eta_C$ and near $\eta\simeq \eta_{\rm min}$. Here, $\eta\simeq \eta_{\rm min}$ is estimated from the minimum of the efficiency PDFs shown in the top figure (vertical cyan line). \label{fig:LDFS}}
\end{figure}

The convergence of work and heat PDFs with $\tau_{\rm obs}$ at low observation times ($\sim 20$ cycles) suggests that the efficiency LDF could be accurately estimated by the  value of the efficiency rate function $-\ln[\rho_{\tau}^{(\tau_{\rm obs})}(\eta)]\,/\,\tau_{\rm obs} $ for $\tau_{\rm obs}$ of the order of the duration of several tenths of cycles. Figure \ref{fig:LDFS} shows the value of $\ln[\rho_{\tau}^{(\tau_{\rm obs})}(\eta)]\,/\,\tau_{\rm obs}$ for different values of $\tau_{\rm obs}$. At observation times corresponding to $10$, $20$ and $30$ cycles our engine attains  a minimum of the efficiency PDF at an off-Carnot value, $\eta_{\rm min}\simeq 2.5\,\eta_C$ as predicted by the theory of stochastic efficiency for asymmetric cycles \cite{gingrich2014efficiency}.  Using an extrapolation technique described in Sec. \ref{sec:method} we obtain an estimation of the efficiency LDF that lies between the efficiency rate function calculated for $20$ and $30$ cycles (black curve in Fig. \ref{fig:LDFS}). 
Note that the entropy production vanishes when  averaged over cycles that perform an efficiency equal to $\eta_{\rm min}$, as does when averaged over cycles with efficiency equal to the Carnot value (Fig. \ref{fig:LDFS} bottom, cf. Fig. 3 in \cite{gingrich2014efficiency}).

\begin{figure}[ht]
\centering
\includegraphics[width=5cm]{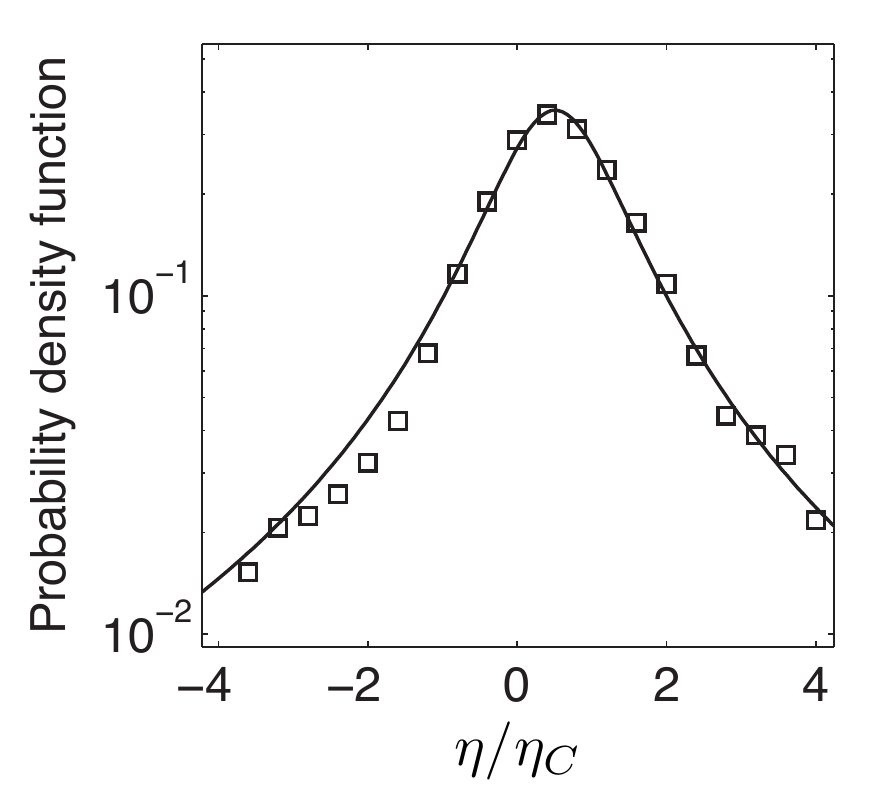}
\caption{\textbf{Distribution of stochastic efficiency $\eta_{\rm loc}$ for the Carnot cycle with $\tau=40\rm ms$}. Experimental value of the one-cycle efficiency distribution obtained with a regular binning of $0.2\eta_C$ (black squares) and fit to a Cauchy distribution (black line). The goodness of the fit is $R^2=0.998$.}
  \label{fig:Cauchy}
\end{figure}

Figure \ref{fig:Cauchy} shows the distribution of the stochastic efficiency $\eta_{\rm loc}$ defined in \cite{sekimoto200carnot} as the ratio between the work extracted and the heat absorbed in one cycle, $\eta_{\rm loc}=\eta_{\tau}^{(1)}$. Our experimental result confirms the theoretical prediction for the one-cycle efficiency distribution, which can be well described by a Cauchy distribution, as predicted for the case of engines with Gaussian heat and work fluctuations \cite{polettini2015efficiency}.

\section{Estimation of the efficiency LDF from finite-time observations}
\label{sec:method}

\begin{figure}
\centering
\includegraphics[width=5cm]{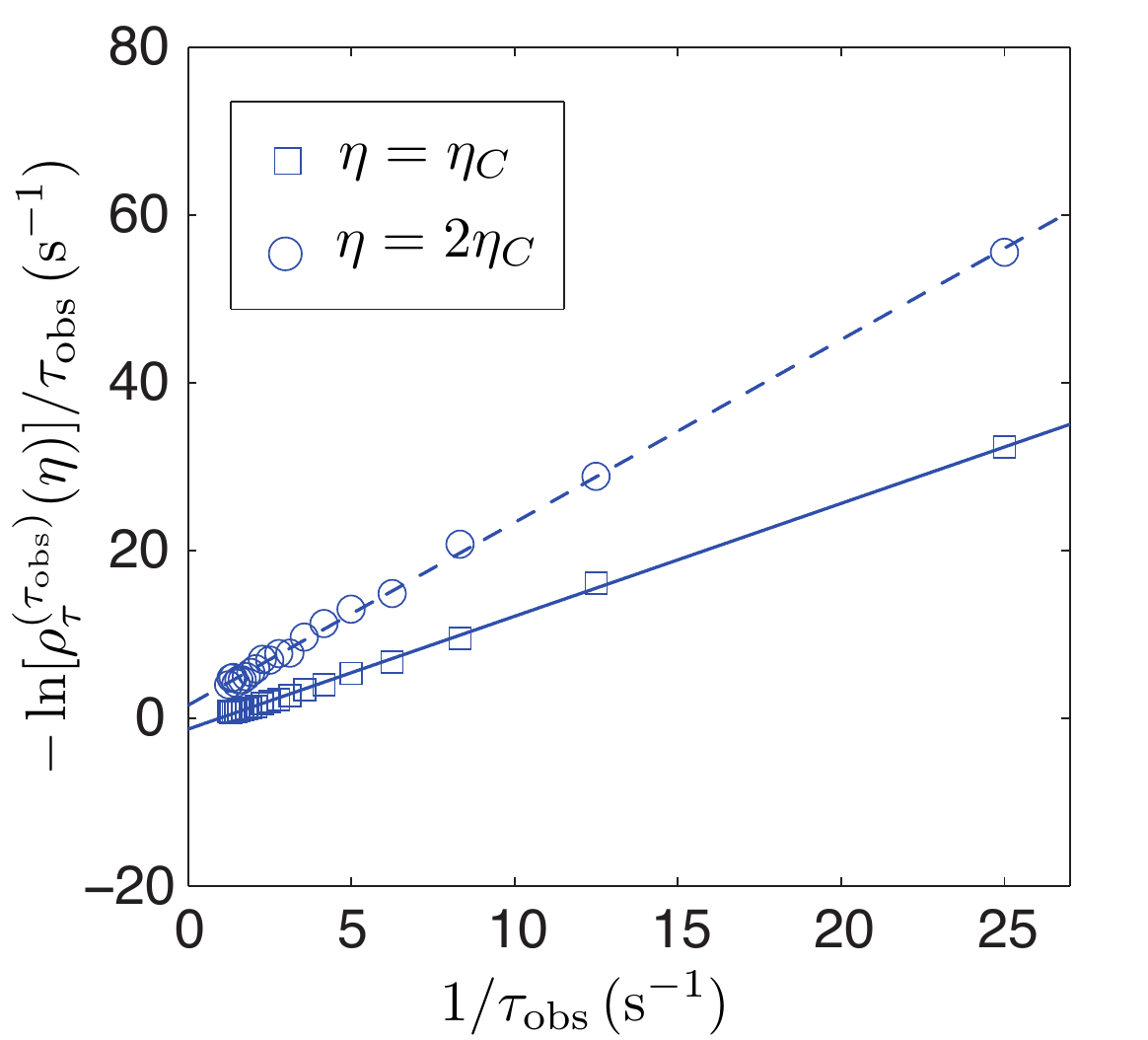}
\caption{\textbf{Estimation of the efficiency LDF from finite $\tau_{\rm obs}$ in the Carnot cycle with $\tau=40\,\rm s$}. Rate function of the efficiency PDF as a function of the inverse observation time for two different values of the efficiency corresponding to $\tau_{\rm obs}$ up to $20$ cycles.  The LDF is estimated as the y-intercept of the linear fit of the rate function vs the inverse observation time (solid and dashed lines). In the data shown here $\mathcal{J}_\tau(2\eta_C) > \mathcal{J}_\tau(\eta_C)$.}
  \label{fig:method}
\end{figure}

\begin{figure}
\centering
\includegraphics[width=5cm]{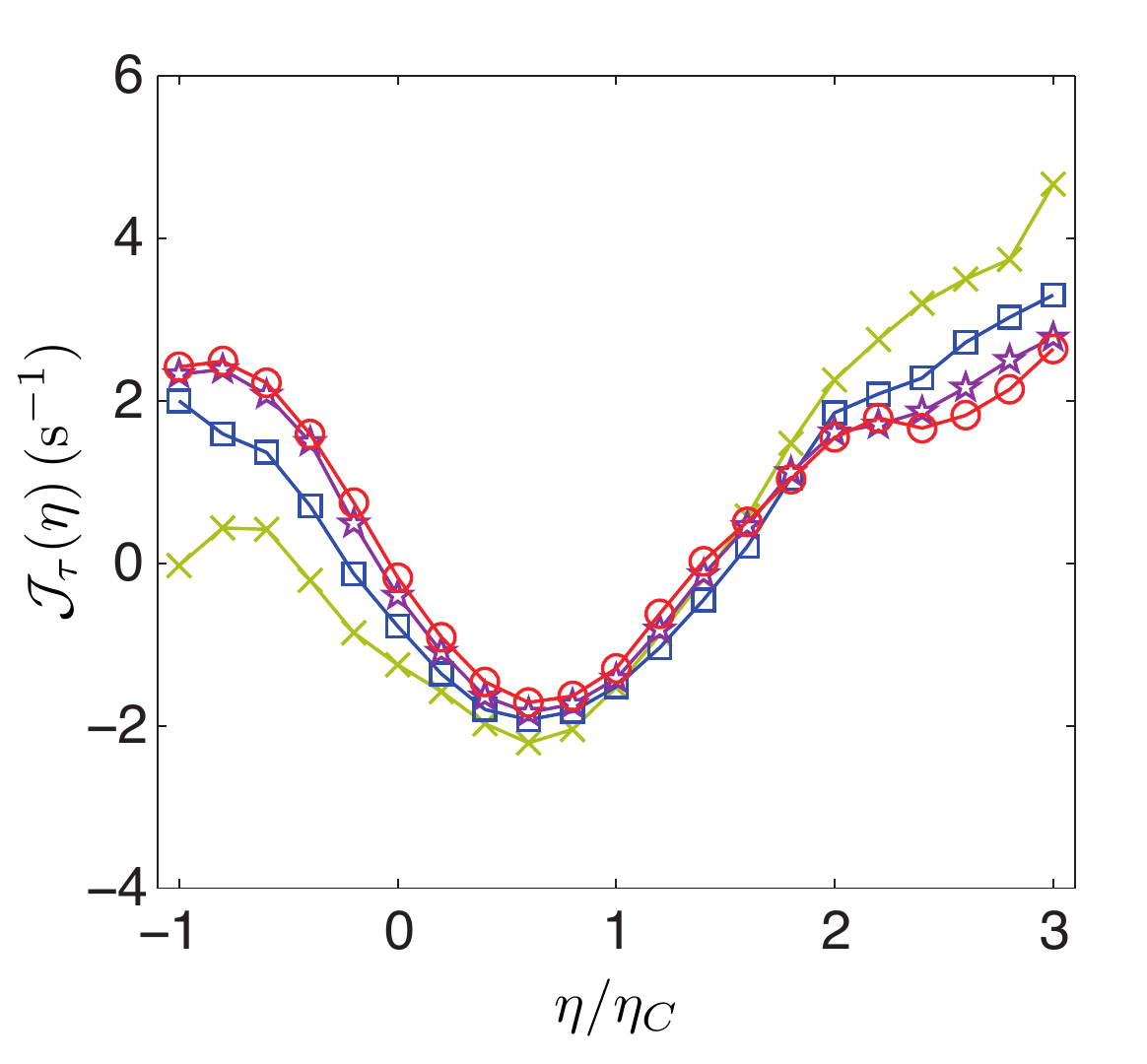}
\caption{\textbf{Efficiency LDF in the Carnot cycle of duration $\tau=40\,\rm ms$}. The data is obtained from linear extrapolation of $-\frac{1}{\tau_{\rm obs}}\ln\rho(\eta_{\tau}^{(\tau_{\rm obs})})$ vs $1/\tau_{\rm obs}$ using the data of efficiency PDFs with $\tau_{\rm obs}$ ranging from $1$ cycle to different number of cycles: $5$ cycles (yellow crosses), $10$ cycles (blue squares), $15$ cycles (magenta stars) and $20$ cycles (red circles). The red curve corresponds with the solid black curve in Fig. \ref{fig:LDFS}.  Solid lines are a guide to the eye. \label{fig:LDFeff}}
\end{figure}

In experimental time series of a finite duration $\tau_{\rm exp}$, the statistics of $\rho(\eta_\tau^{(\tau_{\rm obs})})$ for a large observation time $\tau_{\rm obs}$ is limited. The estimation of the efficiency LDF using Eqs. \eqref{eq:LDF_estimation} and \eqref{eq:LDF_estimation2} is therefore subject to possible statistic shortcomings in the long-time limit. We design an alternative estimator  by extrapolating the rate function $-\ln[\rho(\eta_\tau^{(\tau_{\rm obs})})]/\tau_{\rm obs}$ to $\tau_{\rm obs}=\infty$ from the efficiency PDFs $\rho(\eta_\tau^{(\tau_{\rm obs})})$ for $\tau_{\rm obs}$ small, where the statistics is more robust.  Empirically, we find  the following finite-time correction for the LDF,
\begin{equation}
-\frac{1}{\tau_{\rm obs}}\ln\rho(\eta_{\tau}^{(\tau_{\rm obs})}) = \mathcal{J}_\tau(\eta) + \frac{B}{\tau_{\rm obs}}\quad,
\end{equation}
as shown in Fig. \ref{fig:method} for $\eta=\eta_C$ and $\eta=2\eta_C$. 
As a result, $\mathcal{J}_\tau(\eta)$ can be estimated from the y-intercept of a linear fit of $-\frac{1}{\tau_{\rm obs}}\ln\rho(\eta_{\tau}^{(\tau_{\rm obs})}) $ vs $1/\tau_{\rm obs}$. Figure \ref{fig:LDFeff} shows the value of the estimator of $\mathcal{J}_{\tau}(\eta)$ obtained using this method. In the extrapolation, we use the data of the PDFs $\rho_{\tau}(\eta^{\tau_{\rm obs}})$ for $\tau_{\rm obs}$ ranging from $1$ cycle period to
$\tau_{\rm max}=5\tau$  (yellow crosses), $10\tau$  (blue squares), $15\tau$  (magenta stars) and $20\tau$ (red circles). Our estimator of  $\mathcal{J}_\tau(\eta)$ converges for $\tau_{\rm max}\sim 20\tau$ therefore confirming that the rate function estimation introduced in Sec. \ref{sec:tests} is also an accurate estimator of the efficiency LDF. %that   requires however less computational resources than the extrapolation method introduced here.

\pagebreak

\end{document}